%% file: main.tex
\documentclass[USenglish,oneside,twocolumn]{article}

\usepackage[utf8]{inputenc}
\usepackage[big]{dgruyter}

\newcommand\enquote[1]{`#1'}
\usepackage{booktabs,dingbat}
\usepackage{caption,subcaption}
\newcolumntype{C}{>{\centering\arraybackslash}X}
\usepackage{color}

\DOI{foobar}
\cclogo{\includegraphics{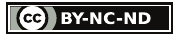}}

\begin{document}

\author*[0]{Konrad Kollnig}
\author[1]{Anastasia Shuba}
\author[2]{Reuben Binns}
\author[2]{Max Van Kleek}
\author[2]{Nigel Shadbolt}

\affil[0]{Department of Computer Science, University of Oxford, E-mail: \mbox{firstname.lastname@cs.ox.ac.uk}}
\affil[1]{Independent Researcher, E-mail: ashuba22@gmail.com}
\affil[2]{Department of Computer Science, University of Oxford, E-mail: firstname.lastname@cs.ox.ac.uk}

\title{\huge Are iPhones Really Better for Privacy? A Comparative Study of iOS and Android Apps}

\runningtitle{Are iPhones Really Better for Privacy? A Comparative Study of iOS and Android Apps}

\begin{abstract}
    {While many studies have looked at privacy properties of the Android and Google Play app ecosystem, comparatively much less is known about iOS and the Apple App Store, the most widely used ecosystem in the US.
	At the same time, there is increasing competition around privacy between these smartphone operating system providers.
    In this paper, we present a study of 24k Android and iOS apps from 2020 along several dimensions relating to user privacy.
    We find that third-party tracking and the sharing of unique user identifiers was widespread in apps from both ecosystems, even in apps aimed at children.
    In the children's category, iOS apps tended to use fewer advertising-related tracking than their Android counterparts, but could more often access children's location.
    Across all studied apps, our study highlights widespread potential violations of US, EU and UK privacy law, including 1) the use of third-party tracking without user consent, 2) the lack of parental consent before sharing personally identifiable information (PII) with third-parties in children's apps, 3) the non-data-minimising configuration of tracking libraries, 4) the sending of personal data to countries without an adequate level of data protection, and 5) the continued absence of transparency around tracking, partly due to design decisions by Apple and Google.
    Overall, we find that neither platform is clearly better than the other for privacy across the dimensions we studied.
    }
\end{abstract}

  \journalname{Proceedings on Privacy Enhancing Technologies}
\DOI{Editor to enter DOI}
  \startpage{1}
  \received{..}
  \revised{..}
  \accepted{..}

  \journalyear{..}
  \journalvolume{..}
  \journalissue{..}

\maketitle

\section{Introduction}
\label{introduction}
\input{1_introduction}

\section{Background}
\label{sec:background}
\input{2_background}

\input{3_method}

\section{Results}
\label{sec:results}
\input{4_results}

\input{5_discussion_conclusions}

\section*{Acknowledgements}

We thank the anonymous reviewers of the Privacy Enhancing Technologies Symposium, our shepherd Arnaud Legout, Ulrik Lyngs, Jun Zhao, Claudine Tinsman and Martin J Kraemer.

Konrad Kollnig was funded by the UK Engineering and Physical Sciences Research Council (EPSRC) under grant number EP/R513295/1.
Max Van Kleek has been supported by the PETRAS National Centre of Excellence for IoT Systems Cybersecurity, which has been funded by the UK EPSRC under grant number EP/S035362/1.
The other authors received no specific grant from any funding agency in the public, commercial, or not-for-profit sectors for this research.

\bibliographystyle{plain}
\bibliography{references}

\end{document}

%% file: 1_introduction.tex
The collection and processing of personal data has become a nearly ubiquitous part of digital life, and is dominated by a small number of powerful technology companies. This is particularly true for smartphones~\cite{playdrone_2014,maps_2019,china_2018,binns_third_2018}, which have a variety of always-on sensors and are carried everywhere, and where just two companies dominate both the operating systems and app distribution channels: Apple and Google.
Previous research on smartphone privacy has focused on one of these companies~--~Google and the Android ecosystem~\cite{playdrone_2014,binns_measuring_2018,china_2018, han_comparing_2013,ren_recon_2016,van_kleek_better_2017,reyes_wont_2018,privacyguard_vpn_2015,nomoads_2018,shuba_nomoats_2020,free_v_paid_2019,okoyomon_ridiculousness_2019}~--~but limited research exists on Apple's iOS and App Store ecosystem~\cite{maps_2019,binns_third_2018}, which has a market share of nearly two thirds in the US~\cite{ios_marketshare1,ios_marketshare2}.
One reason for the limited existence of iOS privacy research has been the lack of publicly available analysis tools, paired with the encryption of iOS apps and the uncertain legality of their decryption.
Knowledge of the app ecosystem is important both so that consumer choice between platforms can be informed on privacy grounds, but moreover for effective regulation~\cite{house_antitrust,doj_apple_google,eu_appstore,competition_and_markets_authority_online_2020,bundeskartellamt_b6-2216_2019} and democratic debate regarding these increasingly important pieces of digital infrastructure.

Apple and Google govern their respective app ecosystems, but pursue different strategies with respect to revenue streams, and the freedoms and responsibilities they grant to app publishers and users. In terms of revenue streams,
both platforms take a share of up to 30\% from all direct revenues created from app sales and in-app purchases, but
differ otherwise~\cite{holzer_mobile_2011}. Apple profits from the sale of iOS devices and does not license iOS to other device manufacturers.
Google's strategy, in contrast, is geared towards the global distribution of Android and Google Play on handsets manufactured by others~\cite{bergvall-kareborn_futures_2013}. Android itself is open-source, but Original Equipment Manufacturers (OEMs) pay a license to distribute the standard Google apps (including the Google Play Store app). A more significant source of revenue for Google is advertising; the parent company of Google, Alphabet, is estimated to have generated \$147bn (80\%) of its 2020 revenue from advertising~\cite{google_ad_revenue}, with more than half the revenue stemming from mobile devices~\cite{google_mobile_share}. This advertising business greatly relies on the collection of data about users, including from mobile devices. More users mean more data, which, in turn, result in more lucrative ads and revenue. While ads often give users access to software for free, the tracking and real-time bidding infrastructure that lie behind them are also known as a threat to individual privacy and can infringe on users' data protection rights~\cite{vines_exploring_2017,chen_following_2016,reardon_50_2019,binns_third_2018}.

As well as differences in revenue streams, the two platforms differ in their approach to the level of freedom granted to app publishers and users. The Google Play Store grants relative freedom. End-users can modify their devices rather easily, and install apps from sources other than Google Play. The underlying operating system, Android, follows an open-source strategy, which has arguably contributed to its success~\cite{bergvall-kareborn_futures_2013,holzer_mobile_2011}. 
The freedom available on Android is not entirely unbridled. The open-source approach does not extend to many Google services on Android, including the Play Services, upon which most apps depend for push notifications and in-app purchases (IAPs), among other essential functionality. Further, Google exerts discretionary control over apps admitted to the official Google Play store, which includes bans on certain types of apps, such as ad blockers. 
While no explicit justification is needed, the ban on ad blockers is based on the claim that such apps may \enquote{interfere with [\ldots] other apps on the device}~\cite{blocker_ban}. However, in general, Google's restrictions have been much more permissive than those exerted by Apple on the iOS App Store, which have a much more stringent set of restrictions (e.g. regarding user privacy) and regularly use manual review to check for compliance, using criteria that are not always clear~\cite{greene_platform_2018}.

These differences in revenue streams and control over app distribution are often cited to explain the alleged differences in the efforts each platform has made to restrict personal data flows and protect user privacy. Of the two, Apple has arguably placed a larger emphasis on privacy, seeking to gain a competitive advantage by appealing to privacy-concerned consumers~\cite{martin_role_2017}. For instance, as early as 2011, Apple started to phase out all permanent device identifiers, in favour of a user-resettable Advertising Identifier (AdId)~--~also called Identifier for Advertisers (IDFA). At their 2019 developer conference, Apple announced a ban on most third-party tracking from children's apps, a particularly vulnerable group of app users. And in 2021, starting with iOS~14.5, Apple requires developers to ask users for permission before accessing the AdId or engaging in advertising-related tracking.
While Google has followed Apple's lead in restricting the use of permanent identifiers, it
currently does not allow Android users to prevent apps from accessing the AdId.

Given the differences between these business models and the greater emphasis on privacy by Apple, it would be reasonable to assume that the iOS ecosystem would be the more privacy-protective in general, in terms of the kinds of data that can be shared, and the extent of third-party sharing. However, little recent empirical research has tested these assumptions in detail, by comparing privacy practices of apps on the two ecosystems at scale. This work fills this gap, by examining the privacy behaviours of apps on the Apple App Store and Google Play, comparing them explicitly, and examining how particular design decisions underlying the two ecosystems might affect user privacy.

\noindent \textbf{Empirical Contributions.}
Given that there are multiple dimensions of privacy, and a corresponding multiplicity in ways to measure it, we adopt a mixture of different indicators and scales to examine each ecosystem along several complementary facets, as follows:\\
\begin{enumerate}
    \item \textit{Code Analysis} of a representative sample of 12k apps from each platform to assess commonly studied privacy metrics (e.g. permissions and presence of tracking libraries) at scale and across platforms.
    \item \textit{Network Traffic Analysis} of the same 12k apps from each platform to study apps' real-world behaviour.
    \item \textit{Company Resolution} to reveal the companies ultimately behind tracking, as well as the jurisdictions within which they reside.
\end{enumerate}

Using the privacy footprints built from our analyses, we find and discuss violations of privacy law and limited compliance with app stores' data collection policies.
We note that while there exist a few other studies that have looked at \textit{security vulnerabilities} in larger numbers of iOS apps~\cite{orikogbo_crios_2016,tang_ios_2020,chen_following_2016}, this present study is the largest study of \textit{privacy aspects} of apps across Android and iOS to date and of privacy in iOS apps since 2013~\cite{agarwal_protectmyprivacy_2013}.
Analysing apps last updated 2018--2020, we study app privacy shortly before Apple's introduction of mandatory opt-ins to tracking in 2021 with iOS 14.5.

\noindent \textbf{Technical Contributions.}
We present a methodology for large-scale and automatic download, privacy analysis, and comparison of apps from the Google Play and Apple App Stores.
So far, no comparable tools have existed in the public domain,
despite such tools being necessary to understand app privacy at large, and to hold the platform gatekeepers to account.
Compared to previous analysis tools for iOS, our approach does not rely on the decryption of apps.
We make our tools and dataset, including the raw app data, publicly available at \url{https://platformcontrol.org/}.

\noindent \textbf{Structure.}
The remainder of this paper is structured as follows. We first summarise the challenges in analysing iOS apps and review related work in Section~\ref{sec:background}.
Next, we introduce our app download and analysis methodology of 12k apps from each app platform in Section~\ref{sec:methodology}.
We then turn to our results from the code and network traffic analysis in Section~\ref{sec:results}, and examine the companies behind tracking.
We discuss limitations in Section~\ref{sec:limitations}, and conclude the paper and outline directions for future work in Section~\ref{sec:conclusions}.

%% file: 2_background.tex
\begin{table*}
\centering
\begin{tabularx}{\textwidth}{l*{15}{C}c}
  & \multicolumn{2}{@{}c@{\hskip0.25in}}{\normalfont \textsf{Android Only}} & \multicolumn{2}{@{}c@{\hskip0.25in}}{\normalfont \textsf{iOS Only}}  & \multicolumn{3}{@{}c@{\hskip0.25in}}{\normalfont \textsf{Android \& iOS}} \\ 
 & Viennot~\cite{playdrone_2014} & Binns~\cite{binns_third_2018} &
Agarwal~\cite{agarwal_protectmyprivacy_2013} & Egele~\cite{pios_2011} & Han~\cite{han_comparing_2013} & Ren~\cite{ren_recon_2016} & This paper \\ \midrule
Publication Year & 2014 & 2018 & 2013 & 2011 & 2013 & 2016 & 2021 \\
Total Apps & 1m & 1m  & 226k & 1.4k & 2.6k & 200 & 24k \\ 
Static Analysis & \checkmark & \checkmark  & x & \checkmark & \checkmark & x & \checkmark \\ 
Dynamic Analysis & x & x & \checkmark & x & x & \checkmark & \checkmark \\
Tracking Libraries   & \checkmark & \checkmark & x & x & \checkmark & x & \checkmark \\
Permissions & x & x & x & x & \checkmark & x & \checkmark \\
PII Usage & x & x & \checkmark & \checkmark & x & \checkmark & \checkmark \\
\end{tabularx}
\caption{Previous papers studying \textit{privacy} properties of iOS and Android apps.
We only include a small subset of important `Android Only' studies.
We do not include papers that focus on security vulnerabilities of apps.
}
\label{tab:literature}
\end{table*}

\subsection{Challenges on iOS}
\label{sec:background-summary}

While many studies have analysed privacy in the Android ecosystem, comparatively much less is known about iOS.
One reason for this lies in the few unique challenges that the Apple ecosystem poses.
First, the closed-source nature of the underlying operating system (iOS), including the use of Apple-only programming languages and compilers, complicates analysis efforts.
Previous work managed to decompile a subset of iOS apps, but no universal decompilation tools exist~\cite{pios_2011,zimmeck_2017}.
Another challenge is Apple's \emph{FairPlay DRM}, which makes accessing and analysing apps' code more difficult than on Android.
Decryption is possible, but relies on access to a physical device and takes time~\cite{orikogbo_crios_2016,chen_following_2016,pios_2011}.
Depending on the jurisdiction, there might also be legal challenges related to the decryption of iOS apps, since this circumvents copyright protections (though arguably not particularly effective ones).
However, there exist exemptions for research purposes in some jurisdictions (e.g. the UK and US), the analysis of which is beyond the scope of this paper.

Apple's use of proprietary technologies and copyright protections acts as a deterrent to developing scalable download and privacy analysis tools for iOS.
No publicly available, scalable tools exist for the Apple App Store (unlike for Google Play)~\cite{maps_2019,binns_third_2018,orikogbo_crios_2016}.
However, such tools are necessary to understand the iOS ecosystem at large, and to hold the platform gatekeepers to account.
This work address this gap by introducing tools and methods for both scalable download and analysis of iOS apps without relying on app decryption (see Section~\ref{sec:code-analysis-trackers}).
This allows us to share our tools publicly, without having to worry about uncertain liability.

\subsection{Related Work}
\label{sec:background-related}

Previous research extensively studied privacy in mobile apps.
Key pieces of literature are summarised in Table~\ref{tab:literature}, and are discussed next.
Two main methods have emerged in the academic literature: dynamic and static analysis.

\emph{Dynamic analysis} observes the run-time behaviour of an app and gathers evidence of sensitive data leaving the device. Early research focused on OS instrumentation, i.e. modifying Android~\cite{enck_taintdroid_2010} or iOS~\cite{agarwal_protectmyprivacy_2013}. With the growing complexity of mobile operating systems, recent work has shifted to analysing network traffic~\cite{privacyguard_vpn_2015,nomoads_2018,free_v_paid_2019,reyes_wont_2018,van_kleek_better_2017,ren_recon_2016,lumen_2018}.
Regardless of the method used, dynamic analysis comes with limitations.
One problem is limited scalability, since every app is executed individually. 
Another issue is that not all privacy-relevant parts of apps may be invoked during analysis, potentially leading to incomplete results.

\emph{Static analysis} dissects apps without execution. Usually, apps are decompiled, and the obtained program code is analysed~\cite{han_comparing_2013,pios_2011}.
The key benefit of static analysis is that it can analyse apps quickly, allowing it to scale to millions of apps~\cite{china_2018,playdrone_2014,binns_third_2018,chen_following_2016}.
However, static analysis~--~unlike dynamic analysis~--~does not allow the observation of real data flows because apps are never actually run.
Programming techniques, such as the use of code obfuscation or native code, can pose further obstacles.
This is especially true for iOS apps, which are harder to decompile and are encrypted by default (see Section \ref{sec:background-summary}).

Table~\ref{tab:literature} evaluates prior work based on the used analysis technique (static vs. dynamic) and on the studied privacy properties: (i) \textit{tracking libraries}, (ii) \textit{permissions}, and (iii) \textit{PII usage}.

\textbf{Tracking Libraries.}
Several studies exist that examine the presence of tracking libraries in apps.
For instance, Viennot et al.~\cite{playdrone_2014} analysed more than 1 million apps from the Google Play Store in 2014, and found that 36\% of analysed apps contained the Google Ads library, 12\% the Facebook SDK, and 10\% Google Analytics.
Similarly, Binns et al.~\cite{binns_third_2018} decompiled and analysed third-party data collection in about 1m Google Play apps in 2018. 
The authors found a strong concentration of data collection with very few tracker companies (`trackers`), with Google and Facebook being most prominent.
Chen et al.~\cite{chen_following_2016} decompiled \textasciitilde1.3m Android and \textasciitilde140k iOS apps, and found potentially malicious libraries in 7\% of Android and 3\% of iOS apps in 2016.

\textbf{Permissions.}
Analysing permission use by apps has a long history in app research~\cite{felt_android_2011,reardon_50_2019,van_kleek_better_2017,jeon_dr_2012,liu_follow_2016,lin_modeling_2014,wijesekera_feasibility_2017,felt_effectiveness_2011,barrera_methodology_2010,han_comparing_2013}.
For instance, Han et al.~\cite{han_comparing_2013} decompiled and analysed 1,300 pairs of iOS and Android apps in 2013.
They found that iOS apps accessed sensitive data
significantly more often than their Android counterparts.
Advertising and analytics libraries accounted for a third of these accesses.
The analysis of permissions only gives a partial picture of apps' privacy practices, since apps tend to request more permissions than necessary~\cite{felt_android_2011}, but may never access the information associated with the permission.
Moreover, some Android apps have even been found circumventing the permissions system~\cite{reardon_50_2019}.

\textbf{PII Usage.} 
There are several approaches to study PII usage in apps.
Some approaches, such as the one taken by Agarwal and Hall~\cite{agarwal_protectmyprivacy_2013} in 2013, examine \textit{access} to sensitive data by intercepting API calls in a jailbroken iOS device.
Since access does not always lead to \textit{transmission}, recent work has shifted to a network-based approach to detect PII \textit{exposure} over the network~\cite{privacyguard_vpn_2015,shuba2016antmonitor,free_v_paid_2019,reyes_wont_2018,van_kleek_better_2017,ren_recon_2016}.
For example, Ren et al.~\cite{ren_recon_2016} developed a VPN server to detect the sharing of PII independent of the mobile operating system in 2016.

\textbf{Our Work.} 
In this paper, we would like to provide an updated study of privacy practices in apps across Android and iOS
at sufficient scale.
Most of the studies discussed above examine either the Android or the iOS ecosystem.
The number of comparative studies is limited, so we seek to address this gap.
Unlike previous work, we would like to analyse iOS apps without relying on app decryption or only traffic analysis, to retrieve rich insights about app privacy at scale through both dynamic and static analysis, and to make our analysis toolchain public at \url{https://platformcontrol.org/} without having to worry about uncertain liability.

%% file: 3_method.tex
\begin{figure*}
    \centering
    \includegraphics[width=0.75\textwidth]{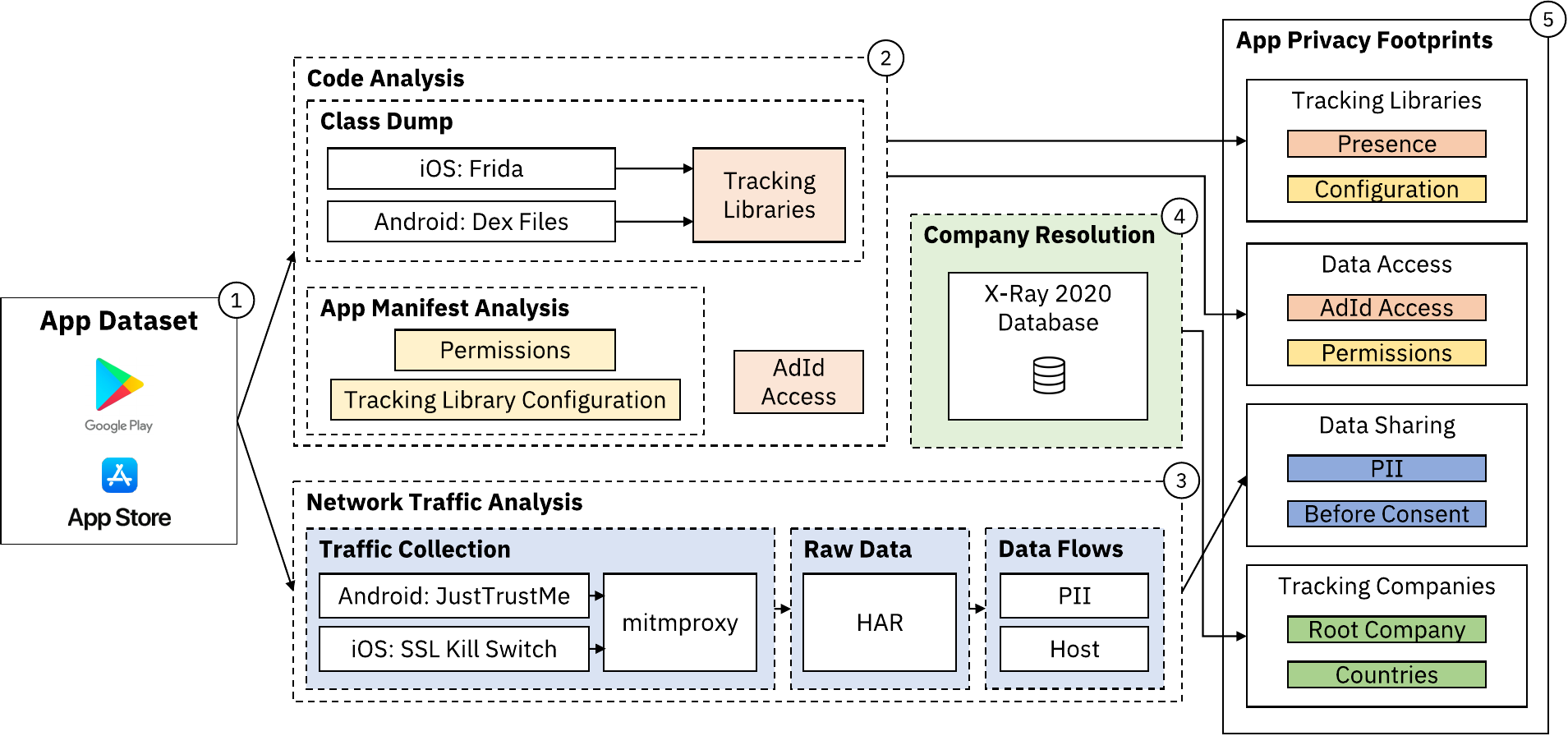}
    \caption{Overview of our analysis methodology (Section \ref{sec:methodology}): First, (1) we select and download 12k apps from the Google Play and Apple App Stores each (Section \ref{sec:downloading}). We then perform a (2) \textbf{Code Analysis} (Section \ref{sec:code-analysis}): (i) we inspect the list of class names (obtained from \texttt{*.dex} files on Android and Frida class dumps on iOS) for known tracking libraries; (ii) we check if apps can access the AdId; and (iii) we analyse the App Manifest to obtain a list of permissions and also to determine the privacy configuration of popular tracking libraries. Third, (3) we 
    conduct a \textbf{Network Traffic Analysis} (Section~\ref{sec:network-analysis}): we disable certificate validation and execute each downloaded app while using \texttt{mitmproxy} to capture network traffic in the \texttt{HAR} format. We analyse the captured traffic for occurrences of PII and contacted host names. Finally, (4) we perform a \textbf{Company Resolution} (Section~\ref{sec:company_resolution}) to obtain a list of companies behind tracking, their owner companies, and the countries of these companies. We use the X-Ray 2020 database for this analysis and resolve the companies behind both the identified tracking libraries in (2) and the contacted hosts in (3). The results of this analysis (Section \ref{sec:results}) are detailed \textbf{App Privacy Footprints} (5) of the downloaded apps, that allow for comparison of privacy characteristics between the two platforms.}
    \label{fig:method_flow}
\end{figure*}

\section{Methodology}
\label{sec:methodology}
In this Section, we describe our analysis methodology, depicted in Figure \ref{fig:method_flow}.
We begin by detailing our app selection and download process in Section \ref{sec:downloading}.
Next, in Section \ref{sec:code-analysis}, we present our method for code analysis, which allows us to extract the following information about each app (without the need to decrypt iOS apps): what tracking libraries are used, how they are configured, which permissions are requested, and whether or not the AdId is accessed. 
Afterwards, in Section \ref{sec:network-analysis}, we describe how we collected decrypted network traffic and analysed it for PII exposure.
Finally, in Section \ref{sec:company_resolution}, we provide details on how we resolved tracking activities (found by both code and network analysis) to the companies behind them and their country of origin.

\subsection{App Dataset and Download}
\label{sec:downloading}

This section details our process for selecting and downloading apps from the Google Play and Apple App Stores (step 1 in Figure~\ref{fig:method_flow}). We also discuss the statistical soundness of our app corpus and our methodology for identifying cross-platform apps within it.
Our code and data are available at \url{https://platformcontrol.org/}.

\textbf{App Selection.}
To select apps, we fed the auto-complete search functionality of the respective app stores with alphanumeric strings of up to three characters to identify popular search terms, similar to previous literature~\cite{playdrone_2014,binns_third_2018}.
Searching for these terms on the app stores then allowed us to identify large numbers of apps, and collect relevant meta information (including title, release date, and time of last update).
We restricted our analysis to apps available in the UK region for both app stores, on the basis that such apps must comply with the General Data Protection Regulation (GDPR)~\cite{gdpr}. Despite the UK's withdrawal from the EU, the GDPR remains applicable in the UK, since it had already been translated into domestic law.
In addition, we only considered apps released or updated in 2018 or later, to focus on apps currently in use.

In total, we identified 568\,745 free apps over 2.5 months between December 2019 and February 2020.
This is before the introduction of Apple's new opt-in mechanism for tracking in 2021.
Our dataset therefore reflects privacy in the app ecosystem shortly before this policy change.
The number of iOS apps ($n=$ 285,680) and Android apps ($n=$ 283,065) was similar.

\textbf{App Download.}
From our selection of 568\,745 apps, we downloaded a random subset of 24,000 apps (12,000 from each platform) for further analysis in this paper.
Our download methodology expands on the App X-Ray project, which is open-source~\cite{xray-code} and has previously enabled the analysis of \textasciitilde1 million Android apps in 2018~\cite{binns_third_2018}.
Adding to this project, we have 1) implemented a scalable download method for the Apple App Store, and 2) restored compatibility with the latest API changes of the Google Play Store to enable the download of Android apps at scale.

The X-Ray project uses the existing Python library \texttt{gplaycli}~\cite{gplaycli} to download Android apps from the Google Play Store.
For the Apple App Store, we used the automation tool \texttt{AutoHotkey}~\cite{autohotkey} to interact directly with Apple iTunes, through its Component Object Model (COM) interface. For each identified iOS app, a purpose-built \texttt{AutoHotkey} script opened the app's download page in the Windows version of iTunes and clicked the \emph{Download} button, so as to download the app, similar to Orikogbo et al.~\cite{orikogbo_crios_2016}.

\textbf{Statistical Extrapolation from Sample.}
In this paper, we are interested in studying trackers, and thus we need to ensure that the results we gather on tracking activities in our app corpus can be extrapolated.
Here, we will argue that our corpus of 24,000 apps is statistically sound when it comes to tracking libraries.
For a description of how we identify tracking libraries, see Section \ref{sec:code-analysis}.
We chose to download more than 10,000 apps for each platform to bring down the margin of the 95\% confidence interval (and thereby, the sampling error in our dataset) for the tracker prevalence $\overline{X_T}$ to less than 2\%, for every studied tracker $T$, assuming an underlying normal distribution due to the law of large numbers:
\begin{equation*}
    \overline{X_T} \sim N(\mu, \sigma^2)
\end{equation*}
Studying a random subset of $100$ or even $1000$ Android apps would not suffice to reach this sampling error margin of $2\%$.
For example, the expected 95\% confidence interval for containing the Facebook SDK was $(19.2\%, \, 37.0\%)$ for a sample of $100$ apps from our dataset, yielding an expected sampling error margin of $17.8\%$.
For a sample of $1,000$ Android apps: $(25.3\%, \, 30.9\%)$, yielding an expected sampling error margin of $5.6\%$. 
However, in our dataset of $12,000$ Android apps, $28.1\%$ of apps contained the Facebook SDK; the $95\%$ confidence interval was $(27.3\%, \, 28.9\%)$.
In conclusion, while we focus on a subset from the overall apps, our results can be extrapolated to the larger dataset, and across all apps on the app stores updated since 2018, with limited error.

Additionally, where appropriate, we conduct permutation tests (using 10,000 permutations) to assess the statistical significance of any quantitative comparisons.
In our tests, we use the difference in mean as our test statistic.
Where we do not find statistical significance ($p>0.05$), we also report 95\% confidence intervals.

\textbf{Identification of Cross-Platform Apps.}
While the majority of our paper analyses the set of 24,000 downloaded apps, Section \ref{sec:cross-platform-apps} examines \textit{cross-platform apps}~--~using a simple similarity algorithm that examined terms from both app titles and app identifiers as follows:
We first tokenised, counted and frequency weighted terms from app titles and app identifiers for all 560k iOS and Android apps using TF-IDF, then computed cosine similarities between pairs of the resulting vectors. Among the 24k downloaded apps, we considered only those apps as cross-platform that had a cosine similarity of at least $95\%$.  This amounted to 13.7\% of downloaded Android apps, and 12.8\% of iOS apps.

\subsection{Code Analysis}
\label{sec:code-analysis}
In this section, we describe how we analyse the apps' code in order to assess the usage and configuration of tracking libraries, access to the AdId, and requested permissions (see step 2 in Figure~\ref{fig:method_flow}).

\subsubsection{Tracking Libraries: Presence}
\label{sec:code-analysis-trackers}

\textbf{Tracking Library Detection.}
We first obtained the class names in Android apps directly from their corresponding \texttt{*.dex} files, while for iOS, we used the \texttt{Frida} dynamic instrumentation toolkit to dump class names from apps.
Note that decryption of iOS apps was not necessary with this \texttt{Frida}-based approach.
We then studied what class names occurred in at least 1\% of Android or iOS apps and are related to tracking, similar to Han et al.~\cite{han_comparing_2013}.
We resolved class names to tracking libraries using various online resources, including the Exodus Privacy project for Android apps~\cite{exodus-stats} and the CocoaPods Master repository for iOS ones~\cite{cocoapods} (containing information on class signatures) as well as trackers' online resources (documentation and GitHub repositories).
We identified a total of $40$ tracking libraries of interest, all of which existed for both Android and iOS, with the exception of Google's Play Services (which was present for Android only) and Apple's SKAdNetwork (which was present in iOS only).

\textbf{Impact of Obfuscation.} While some very popular apps may use code obfuscation to hide their tracking activities intentionally, we found that it had little effect on our overall analysis.
By default, iOS apps do not apply any obfuscation to the class names, and developers are well known to be subject to a `default bias' (i.e. not to change default settings) in the literature~\cite{acquisti_privacy_2015,chitkara_does_2017,hadar_privacy_2018,mhaidli_we_2019}.
As for Android, we found similar results by checking against the obfuscation-resilient LibRadar++ library~\cite{libradar,china_2018}.
An important reason for this result is that, while tracking libraries may obfuscate their internal code, obfuscating user-facing APIs is difficult~\cite{chitkara_does_2017}. 
Further, many tracking libraries use inter-app communication and cannot easily obfuscate their communication endpoints~\cite{reardon_50_2019}.
We do not use LibRadar++ for our overall analysis, since it is closed-source, no longer maintained, has an outdated database of library signatures (last updated in 2018) and struggled with different library configurations (for instance, Google Firebase is a set of different libraries, including advertising and analytics components that share some of the same code, but LibRadar++ summarised all these components as \texttt{com.google.firebase}).
We also tried LibScout~\cite{derr:ccs16,derr:ccs17} for library detection, but found that it also missed essential libraries and took long to execute.

\subsubsection{AdId Access}
\label{sec:code-analysis-adid}
The AdId is a unique identifier that exists on both iOS and Android. It allows advertisers to show more relevant ads for users (e.g.~by avoiding showing the same advert twice in two different apps), but it can also be used to create fine-grained profiles about app users~--~something many users may not expect and that can lead to potential violations of data protection law~\cite{kollnig_2021,reyes_wont_2018,nguyen_share_first_consent_2021}.
The AdId is also the only cross-app identifier that may be used for advertising on Android, but might additionally be used for analytics~\cite{google_ad_id}. That is why AdId access might be an upper bound on the use of any form of analytics on Android (including personalised ads); there are no incentives not to use the AdId for these purposes.
While users can theoretically reset the AdId, most do not know how or why to do so~\cite{acquisti_privacy_2015,acquisti_nudging_2009}.
Starting with iOS 14.5 in 2021, the operating system has switched from an opt-out to an opt-in mechanism to apps' use of the AdId; in our study, we will assess privacy in the app ecosystem immediately before this policy change.
We detected potential access to the AdId by checking for the presence of the \texttt{AdSupport} class and the system interface \texttt{IAdvertisingIdService} in the app code on iOS and Android, respectively.

\subsubsection{App Manifest Analysis}
\label{sec:code-analysis-manifest}
\textbf{Permissions.}
Permissions form an important part of the security models of Android and iOS as they protect sensitive information on the device.
We extract the permissions used by the apps in our dataset by parsing the app manifest files.
At the time of data collection, Android defined a total of $167$ permissions, $30$ of which were designated as \textit{dangerous permissions} by Google and require user opt-in at run-time.
Similarly, Apple defined $22$ permissions that needed to be disclosed in the app manifest. All of these require user opt-in.
We only include permissions defined by the Android or iOS operating system, and exclude custom permissions by other vendors (used by some Android apps).
While the targeted OS version can affect what permissions apps can request, only a few new permissions have been added in 2018--2020 and we did not consider this aspect further;
in our results, none of the top 10 permissions on either platform has been added in the period 2018--2020, so this should not significantly affect our reported descriptive statistics.

In addition to reporting statistics on general permissions usage, we further focus on the ones that both Apple and Google agree to be particularly dangerous and need user opt-in.
There are a total of 7 such \textit{cross-platform permissions} that exist on both platforms: Bluetooth, Calendar, Camera, Contacts, Location, Microphone, and Motion.
This total number is small compared to the overall number, since we excluded and summarised some permissions to overcome the different functionality and granularity in permissions across the platforms.
For instance,
Android discriminates between read and write permissions for contacts and calendar, but we have summarised them as \emph{Contacts} and \emph{Calendar}, respectively.

\textbf{Tracking Library Configuration.} Many tracking libraries allow developers to restrict data collection using settings in the app manifest, e.g. to disable the collection of unique identifiers or the automatic SDK initialisation at first app start.
This can help setting up tracking libraries in a legally compliant manner.
For the Facebook SDK, these options were only added after public backlash over the mandatory and automatic sharing of personal information at the first app start, potentially violating EU and UK privacy law~\cite{privacy_international_how_2018}.
We focus on the privacy settings provided by some of the most prominent tracking libraries: Google AdMob, Facebook, and Google Firebase.

\subsection{Network Traffic Analysis}
\label{sec:network-analysis}
In this section, we discuss our network traffic analysis process (step 3 in Figure~\ref{fig:method_flow}).

\textbf{App Execution and Network Traffic Capture.}
We opened every app automatically on a real device~--~a Google Nexus 5 running Android 7 and an iPhone SE 1st Gen with iOS 14.2~--~for 30 seconds without user interaction.
We captured apps' network traffic using \texttt{mitmdump} to study apps' data sharing with tracking domains.
Tracking libraries are usually initialised at the first app start and often without user consent~\cite{kollnig_2021, reyes_wont_2018,nguyen_share_first_consent_2021}, which we aimed to detect with this approach.
We did not perform any further automated actions to the studied apps, since there did not exist established approaches (like the UI Exerciser Monkey on Android) to instrument \textit{arbitrary} iOS apps.

\textbf{Device Configuration.}
We disabled certificate validation using \texttt{JustTrustMe} on Android and \texttt{SSL Kill Switch 2} on iOS, after gaining system-level access on both devices (known as `root' on Android and `jailbreak' on iOS), in order to read apps' HTTPS traffic.
We would have liked to use a more recent version of Android, but we found that disabling certificate validation was unstable on the latest versions of Android.
We note that several identifiers are inaccessible as of Android 10, but 
apps should behave similarly otherwise.
We uninstalled or deactivated pre-installed apps, and were not logged into an Apple or Google account.
On both phones, we did not opt-out from ad personalisation from the system settings, thereby assuming implicit user opt-in to apps' use of the AdId.

\textbf{PII Analysis.}
To analyse the sharing of PII and other personal data, we conducted a case-insensitive search on the network traffic for the following identifiers as well as common transformations thereof (MD5, SHA-1, SHA-256, SHA-512, URL-Encoding): Advertising ID, Android Serial Number, Android ID, phone and model name, and WiFi MAC Address.
We have refrained from analysing further PII, such as location, contacts or calendar, due to lack of instrumentation methods for iOS to get around opt-in permission requests. We also assembled a list of contacted host names. 

\begin{figure}
	\centering
	\includegraphics[width=0.6\columnwidth]{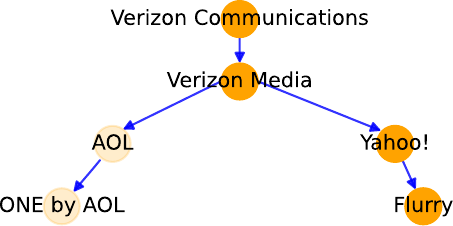}
	\caption{Company structure of Verizon's tracking business. Over recent years, Verizon has sold many of its subsidiaries (including Flickr and Tumblr) and has integrated AOL and its advertising business `ONE by AOL' (both in light orange) into Verizon Media.}
	\label{fig:verizon}
\end{figure}

\subsection{Company Resolution}
\label{sec:company_resolution}
In this Section, we explore
which companies are ultimately behind tracking, and in which jurisdiction these are based in (step 4 in Figure~\ref{fig:method_flow}).
We combine the insights from both the studied tracking libraries (Section~\ref{sec:code-analysis-trackers}), as well as all tracking domains observed in at least 0.5\% of apps' network traffic (Section~\ref{sec:network-analysis}).
Knowledge about the company behind tracking, including its jurisdiction, is essential for the legal assessment of tracking practices.
For this purpose, in 2017, Binns et al. created the X-Ray database~--~a database of known tracker companies, their tracking domains, and their company hierarchies~\cite{binns_measuring_2018}.
We updated this database to mid 2020, to understand the company relations behind tracking, as well as detect what contacted hosts are known to be used for tracking.
This update was necessary since the tracking ecosystem continuously changes.
For instance,
the investment company Blackstone purchased the mobile advertising company Vungle in 2019.
Verizon sold many of its subsidiaries (including Flickr and Tumblr) over recent years and has integrated its subsidiary AOL and its advertising business (`ONE') into Verizon Media (see Figure~\ref{fig:verizon}).
For the update, we followed the protocols of the previous study.
Specifically, for every company in the database,
we checked what parent companies it might have, using WHOIS registration records, Wikipedia, Google, Crunchbase, OpenCorporates, and other public company information.
Our analysis of tracker libraries and domains identified $24.4\%$ additional companies (from $578$ to $754$ companies), and increased the database size by $28.1\%$.
We call the resulting dataset \texttt{X-Ray 2020}, which we made publicly available at \url{https://platformcontrol.org/}.

%% file: 4_results.tex
In this section, we present our findings from analysing 24,000 apps from iOS and Android (step 5 in Figure~\ref{fig:method_flow}).
We analysed 0.86~TB of downloaded apps, and collected 24.2~GB of data in apps' network traffic.
Installing and instrumentation failed for 124 Android and 36 iOS apps; we have excluded these apps from our subsequent analysis.

First, we focus on the tracking libraries found from the code analysis and whether or not they were configured for data minimisation (Section~\ref{sec:static_tracking}).
Next, in Section~\ref{sec:data_access}, we analyse potential data access of apps, by examining their permissions and their access to the AdId.
Following up, in Section~\ref{sec:data_sharing}, we report on the actual data sharing of apps before consent is provided, as well as the observed exposure of PII in network traffic.
Afterwards, we explore the complex network of companies behind tracking and their jurisdictions (Section~\ref{sec:tracking_companies}).
Lastly, we focus on cross-platform (Section~\ref{sec:cross-platform-apps}) and children's apps (Section~\ref{sec:kids-apps}).
Cross-platform apps have received attention in previous studies, but might feature different privacy properties than apps on the ecosystem overall.
Children's apps must adhere to stricter privacy rules, arising both from legal requirements (e.g. COPPA~\cite{coppa} in the US and GDPR in the EU) and the policies of the app store providers.

\begin{figure*}
    \begin{subfigure}{0.49\linewidth}
	    \centering
	    \includegraphics[width=\linewidth]{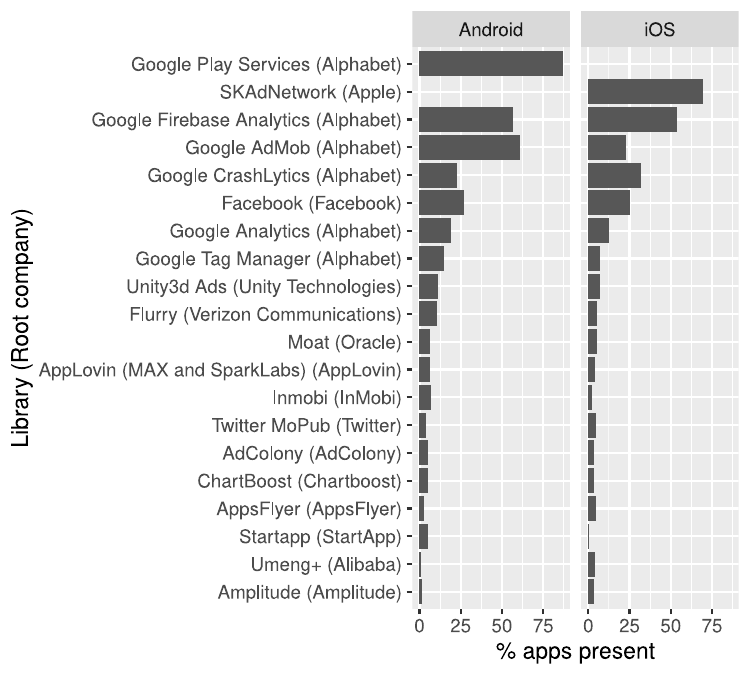}
	    \centering
	    \begin{tabular}{lrrrrrr}
			& Median
			& Mean
			& Q1
			& Q3
			& Count $>10$
			& None \\
			\midrule
			Android & 3 & 3.8 & 2 & 5 & 3.73\% & 11.27\% \\
			iOS & 3 & 3.1 & 1 & 4 & 3.13\% & 20.65\% \\
		\end{tabular}
		\caption{Top tracking libraries in app code.}
		\label{fig:tracker_libraries}
	\end{subfigure}
	\begin{subfigure}{0.49\linewidth}
	    \centering
	    \includegraphics[width=\linewidth]{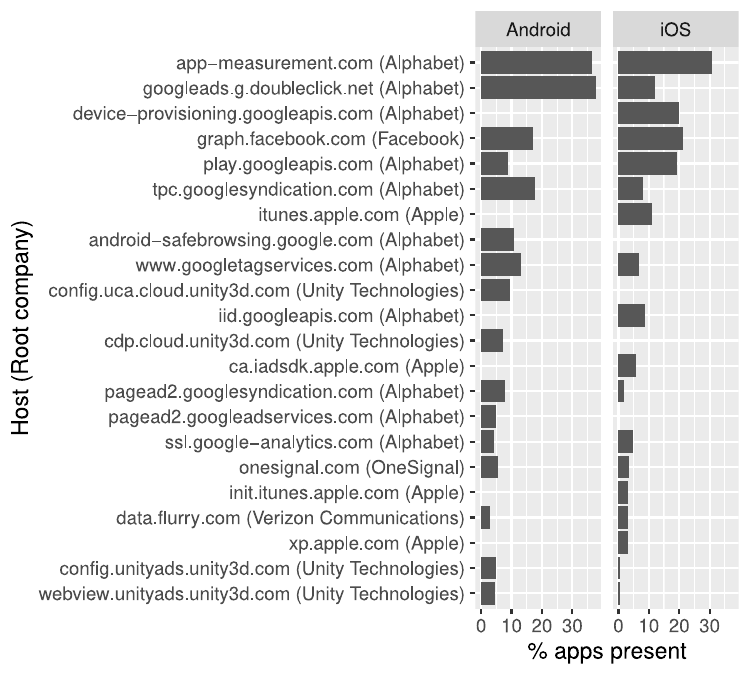}
	    \begin{tabular}{lrrrrrr}
			& Median
			& Mean
			& Q1
			& Q3
			& Count $>10$
			& None \\
			\midrule
			Android & 2 & 2.7 & 1 & 4 & 1.99\% & 18.56\% \\
			iOS & 2 & 2.4 & 0 & 4 & 1.35\% & 31.54\% \\
		\end{tabular}
		\caption{Top tracking hosts contacted at first app start.}
		\label{fig:tracker_hosts}
	\end{subfigure}
	\caption{Third-party libraries and contacted tracking domains of apps, as well as the companies owning them (in brackets). Shown are the top 15 tracking libraries and domains from each platform.}~\label{fig:apps_trackers}
\end{figure*}

\subsection{Tracking Libraries}
\label{sec:static_tracking}

\subsubsection{Presence}
Apps from both platforms widely use tracking libraries
(see Figure~\ref{fig:tracker_libraries}).
The median number of tracking libraries included in an app was 3 on both platforms.
3.73\% of Android apps contained more than 10, compared to 3.13\% on iOS. 88.73\% contained at least one on Android, and 79.35\% on iOS.

The most prominent tracking library on Android is the Google Play Services (in 87.3.\% of apps)~--~a technology that is ultimately owned by Google's parent company Alphabet.
This library provides essential services on Android devices, but is also used for advertising and analytics purposes.
The most prominent library on iOS is the SKAdNetwork library (in 69.6\% of apps). While part of Apple's privacy-preserving advertising attribution system, this library discloses information about what ads a user clicked on to Apple, from which Apple could (theoretically) build user profiles for its own advertising system.
Google's advertising library (`AdMob') ranks second on Android, and occurs in 61.7\% of apps.
One factor driving the adoption of this library on Android might be that it not only helps developers show ads, but also provides easy access to the AdId (although developers could also implement this manually).
However, this dual use of the tracking library might increase Google's reach over the mobile advertising system, by incentivising the use of AdMob.
Google Firebase is the second most popular tracking library on iOS, occurring in 53.9\% of apps, as compared to 57.6\% on Android.
Facebook (recently renamed \enquote{Meta}), the second largest tracker company, has a far smaller reach than Google, and is only present in 28.0\% of apps on Android and 25.5\% on iOS.
Few tracking services are more popular on iOS:
Google Crashlytics occurred in 31.8\% of apps, and 23.8\% on Android.
MoPub, a Twitter-owned advertising service, was present in 4.71\% of iOS apps, and 4.25\% on Android.
Overall, tracking services are widespread on both ecosystems, but slightly more so on Android, likely in part due to Google's dual role as a dominant advertising company and platform gatekeeper on Android.
However, Google also has a significant presence on iOS, highlighting its dominance in both smartphone ecosystems.

We note that certain libraries have sub-components which can be loaded individually and have different consequences on privacy.
For instance, both Google Play Services and Google Firebase bundle a range of different services, from which developers can choose.
Further, certain libraries provide configuration options that also affect privacy.
While we do not consider all the sub-components of libraries in this study, we do analyse the libraries' configurations, as discussed in the next Section.

\subsubsection{Configuration for Data Minimisation}
\label{sec:static_tracking_config}
Only a small fraction of apps made use of data-minimising SDK settings in their manifest files, e.g. to retrieve user consent before sharing data with trackers.
At the same time, `data minimisation' is one of the key principles of GDPR, as laid out in Article 5, and user opt-in is required prior to app tracking in the EU and UK~\cite{kollnig_2021}.
However, we found that the vast majority of developers did not change trackers' \textit{default options} that might lead to more data sharing than necessary.

Among the apps that used Google AdMob, 2.2\% of apps on iOS and 0.8\% on Android chose to delay data collection.
Among the apps using the Facebook SDK, less than 5\% (2.3\% on Android, 4.6\% on iOS) had delayed the sending of app events, less than 1\% (0.4\% on Android, 0.9\% on iOS) had delayed the SDK initialisation, and less than 4\% had disabled the collection of the AdId (0.9\% on Android, 3.0\% on iOS).
Among apps using Google Firebase,
0.5\% had permanently deactivated analytics on Android and 0.4\% on iOS, 1.2\% had disabled the collection of the AdId on Android and 0.1\% on iOS, and 1.2\% had delayed the Firebase data collection on Android and 0.5\% on iOS.

\subsection{Data Access}
\label{sec:data_access}

\subsubsection{AdId Access}
\label{sec:static_adid}
Potential access to the AdId was more widespread among Android apps than iOS ones.
Among the studied apps, 86.1\% of Android apps could access the AdId, 42.7\% on iOS, allowing them to track individuals across apps easily.

Advertising and AdId access were often linked. Of those apps with Google AdMob, 100\% on iOS and 99.6\% on Android had access to the AdId. We had similar results for the next most popular advertising services: of apps with Unity3d Ads, more than $99\%$ of apps accessed the AdID; similarly for Moat ($100\%$ of apps), and Inmobi (more than $94\%$ of apps).
Conversely, about 71.3\% of Android apps and 53.4\% of iOS apps with AdId access used Google AdMob, but less than $20\%$ used Unity3d Ads, Moat or Inmobi.
If we assume, for the sake of argument, that an app shows personalised ads if and only if it has AdId access (because there is hardly any reason for apps not use the AdId for personalised ads), this suggests that Google AdMob was present in the majority of apps with personalised ads.
This points to a high \emph{market concentration} towards Google in the digital advertising market~--~which is coming under increasing scrutiny by the competition regulators and policy makers~\cite{competition_and_markets_authority_online_2020,bundeskartellamt_google}.

One reason for the differences in AdId access might be the restrictions set by the platforms themselves. Apple is taking steps against the use of the AdId, which is often linked to advertising.
On submission to the Apple App Store, app publishers have long had to declare that their app only uses the AdId for certain, specific reasons related to advertising.
Additionally, Apple allows iOS users to prevent all apps from accessing the AdId, and even asks for explicit opt-in as of iOS 14.5.
Google does not currently allow users to prevent apps from accessing the AdId on Android.

\subsubsection{Permissions}
\label{sec:static_permissions}

\begin{figure*}[t]
\centering
\begin{subfigure}{0.65\linewidth}
\resizebox{\linewidth}{!}{%
\begin{tabular}{lrc|lrc}
Android Permission                          & Apps (\%) & Opt-in? & iOS Permission                  & Apps (\%) & Opt-in? \\ \midrule
INTERNET                 & 98.7       & x      & PhotoLibrary               & 58.0       & \checkmark     \\
ACCESS\_NETWORK\_STATE   & 94.5       & x      & Camera                     & 56.3       & \checkmark     \\
WAKE\_LOCK               & 70.0       & x      & LocationWhenInUse          & 47.7       & \checkmark     \\
WRITE\_EXTERNAL\_STORAGE & 63.4       & \checkmark     & LocationAlways             & 31.4       & \checkmark     \\
READ\_EXTERNAL\_STORAGE  & 41.4       & \checkmark     & PhotoLibraryAdd            & 27.4       & \checkmark     \\
ACCESS\_WIFI\_STATE      & 40.0       & x      & Microphone                 & 26.2       & \checkmark     \\
VIBRATE                  & 35.8       & x      & Calendars                  & 25.2       & \checkmark     \\
RECEIVE\_BOOT\_COMPLETED & 26.5       & x      & LocationAlwaysAndWhenInUse & 16.8       & \checkmark     \\
ACCESS\_FINE\_LOCATION   & 26.1       & \checkmark     & BluetoothPeripheral        & 16.4       & \checkmark     \\
ACCESS\_COARSE\_LOCATION & 24.8       & \checkmark     & Contacts                   & 16.1       & \checkmark     \\
READ\_PHONE\_STATE       & 21.5       & \checkmark     & Motion                     & 8.0        & \checkmark     \\
CAMERA                   & 21.4       & \checkmark     & Location                   & 7.5        & \checkmark     \\
FOREGROUND\_SERVICE      & 12.1       & x      & AppleMusic                 & 7.1        & \checkmark     \\
GET\_ACCOUNTS            & 10.1       & \checkmark     & BluetoothAlways            & 6.8        & \checkmark     \\
RECORD\_AUDIO            & 9.7        & \checkmark     & FaceID                     & 6.1        & \checkmark     \\
\end{tabular}}
\caption{Most common permissions on iOS and Android.}
\label{fig:permissions_all}
\end{subfigure}
\begin{subfigure}{0.24\linewidth}
    \includegraphics[width=0.81\linewidth]{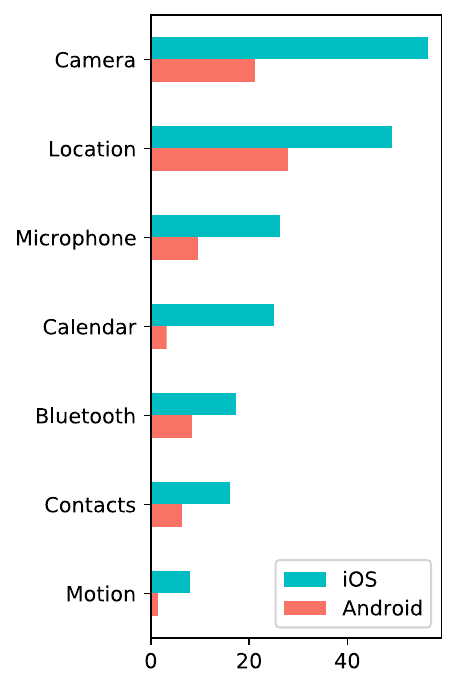}
    \caption{Percentage of apps requesting opt-in permissions. iOS apps consistently included more than Android.}
    \label{fig:permissions_cross_platform}
\end{subfigure}
\caption{Top permissions on Android and iOS. All permissions on iOS require opt-in, only `dangerous' ones on Android.}
\end{figure*}

\textbf{Most Prevalent Permissions.}
Figure~\ref{fig:permissions_all} shows the most prevalent permissions on both platforms, and whether these require opt-in.
The most common permissions on Android are \texttt{INTERNET} and \texttt{ACCESS\_NETWORK\_STATE}, both requested by more than $90\%$ of apps and related to Internet access.
A similar permission does not exist on iOS.
The most common `dangerous' permissions (requiring user opt-in) on Android are related to storing and reading information on the external storage, \texttt{WRITE\_EXTERNAL\_STORAGE} and \texttt{READ\_EXTERNAL\_STORAGE}. Such external storage exists on iOS as well, but apps access it through a system-provided `Files' dialog.
\texttt{PhotoLibrary} (for photo access) is the most common permission on iOS.
Although a similar permission (\texttt{CAMERA}) exists on Android, apps do not have to request it, but can rather invoke the camera application on the phone to take a picture directly.
This potentially explains some of the differences in the number of camera-related permission requests between Android and iOS.
The iOS \texttt{PhotoLibrary} permission was an outlier from the overall observation that iOS apps needed more permissions and was as prevalent (about $60\%$ of apps) as the \texttt{WRITE\_EXTERNAL\_STORAGE} permission on Android, possibly because the most common usage of file access is processing photos (e.g. in social media or photography apps).
Access to external storage can be a privacy risk since it can enable unexpected data exposure and tracking across apps~\cite{reardon_50_2019}.
Because of this, Google has been restricting access to external storage ever more with recent versions of Android and Apple has never allowed direct access to file storage.

\textbf{Cross-Platform Permissions.}
All cross-platform permissions were more common on iOS than on Android, see Figure~\ref{fig:permissions_cross_platform}. The most common were Camera and Location.
Both were included by about 50\% of iOS apps (Camera 56.3\%, Location 49.2\%), and less than a third of Android apps (Camera 21.2\%, Location 28.0\%).
iOS apps also accessed the Calendar and Contacts permissions more often than Android apps (25.2\% vs. 3.2\% for Calendar; 16.1\% vs. 6.4\% for Contacts). Note that Android differentiates between read and write access for the Contacts and Calendar permission.
The studied Android apps with Calendar access usually had both read (95.0\%) and write (94.5\%) access.
Of those with Contacts access, 97.6\% had read and 47.1\% write access, underlining the potential value of separating read and write permissions.
Motion was the least common cross-platform permission, present in 8.0\% of iOS apps and 1.4\% of Android apps.

To seek further explanations on why iOS apps request camera and location access more frequently than Android apps, we first checked the categories of the apps in our dataset. 
Our intuition was that iOS apps would more frequently fall into categories related to photography and navigation, but that was not the case.
Next, we checked the required permissions of the top 15 tracking libraries. However, we did not find any differences, except for AdColony, which requests different permissions on Android and iOS, but has a relatively small market share.
Finally, we measured how many apps mentioned the terms \enquote{photo} or \enquote{camera} in their description on the respective app store.
Including only those descriptions that were in English (according to the \texttt{langdetect} Python library~\cite{langdetect}), 14.0\% of apps on Android and 11.8\% of apps on iOS mentioned either term. However, the median length of descriptions in English was substantially higher on Android (1032 characters) compared to iOS (761 characters) and only 72.7\% of iOS app descriptions were actually in English (81.6\% on Android), making it difficult to interpret these observations.

\textbf{Summary.}
Android has many permissions that have no equivalent on iOS, and thus Android apps can \textit{appear} to be more privileged than their iOS counterparts, but on closer examination, they are simply asking for permissions to access resources which are not restricted on iOS (e.g.~Internet access and network state).
Further, iOS apps showed substantially higher levels of cross-platform permissions that both Apple and Google deem as particularly dangerous and require user opt-in. This can be a reason for concern. Once a permission is granted, an app can usually access sensitive data anytime without the user's knowledge.

From our analysis, it does not seem like the distribution of apps on the app stores, or the different permission requirements of tracking libraries on either platform are the main drivers behind the observed differences in permission use.
Instead, there are a range of architectural differences between the platforms that might lead to increased use of opt-in permissions on iOS.
One important factor might be that Android allows for deeper integration between apps, through its powerful \emph{intent system}. Android apps can call specific functionality of other apps, and listen for return values. 
In the past, Android apps have also been observed to use side channels to circumvent the permission system~\cite{reardon_50_2019}, which underlines the potentially deep integration between Android apps.
By contrast, iOS only allows for very limited cross-app communication.
This might mean that a higher number of dangerous cross-platform permissions on iOS might actually be positive for privacy, since it reflects higher encapsulation of apps.

There has also been a wealth of research into Android's permission system in the past, and much less so on iOS; this, in conjunction with disclosures of permissions on the Google Play Store (but traditionally not so on the App Store), might have made Android developers more cautious about declaring permissions~--~particularly those that require explicit opt-in.

In sum, there are a variety of aspects~--~including differences in software architecture, developer attitudes and practices, and socioeconomics of end-users~--~that drive permission use on either platform. We leave it for future work to disentangle these aspects further.

\subsection{Data Sharing}
\label{sec:data_sharing}

\subsubsection{Before Consent}
\label{sec:data_sharing_consent}
We now turn to the data sharing in apps' network traffic, before any user interaction.
While in this section we do not analyse what personal data is shared, tracker companies necessarily receive the user's IP address from any connections, which itself can classify as personal data under EU law~\cite{ip_addresses} and can be used for tracking purposes~\cite{att_caid1}.
Our results are shown in Figure~\ref{fig:tracker_hosts}.

The average app on both platforms contacted similar numbers of tracking domains (2.7 on Android, and 2.4 on iOS).
18.6\% of Android apps and 31.5\% of iOS apps did not contact any tracking domains at the app start.
The most popular domain (\texttt{googleads.g.doubleclick.net}) on Android was related to Google's advertising business~--~contacted by 37.6\% of Android apps, and 11.9\% on iOS.
The most popular domain on iOS was related to Google's analytics services (\texttt{app-measurement.com})~--~contacted by 30.7\% of apps on iOS, and 36.4\% on Android.
Facebook services were contacted by more iOS apps (21.2\%) than on Android (17.2\%).
Some iOS apps additionally exchange data about app installs (\texttt{itunes.apple.com}) and ad attribution (\texttt{ca.iadsdk.apple.com}) with Apple.
These services are unique to the Apple ecosystem, and do not exist on Android.
As observed in the previous section, advertising services seem more popular on Android than on iOS, by a factor of roughly 2 (e.g. \texttt{*.doubleclick.net}, \texttt{*.googlesyndication.com}, \texttt{unityads.unity3d.com}).

\textbf{Widespread Tracking without Legally Required Consent.}
Overall, we find that data sharing with tracker companies before any user interaction is common on both platforms.
However, EU and UK law requires user consent before third-party tracking can take place~\cite{kollnig_2021}.
This suggests potentially widespread violations of applicable data protection law (in 81.44\% of Android apps, and 68.46\% of iOS apps).
While most of this data sharing can be attributed to Google,
other companies, such as Facebook and Unity, also receive data for tracking purposes.
Moreover, tracking by Google also happens widely on iOS where, unlike on Android, a user would not have given consent as part of the device set-up process.

\subsubsection{PII Exposure}
\label{sec:data_sharing_pii}
We found that more Android apps shared the AdId over the Internet (55.4\% on Android, and 31.0\% on iOS).
The reduced sharing of the AdId on iOS might be related to the reduced prominence of AdId access in iOS apps as found in our static analysis, and the stricter policies by Apple regarding AdId use (see Section~\ref{sec:static_permissions}).
85.1\% of Android and 61.4\% of iOS apps shared the model and phone name over the Internet, which can be used as part of device fingerprinting.

Android apps also shared other system identifiers, including the Android ID (18.2\% of apps), the IMEI (1.3\% of apps), the Serial number (1.1\% of apps) and the WiFi MAC Address (0.6\% of apps).
We note that these identifiers are no longer accessible as of Android 10.
We did not find equivalent identifiers in iOS network traffic; iOS has long deprecated access to permanent identifiers (UDID with iOS 6 in 2012 and MAC Address with iOS 7 in 2013).

\begin{figure}
	    \centering
	    \includegraphics[width=\linewidth]{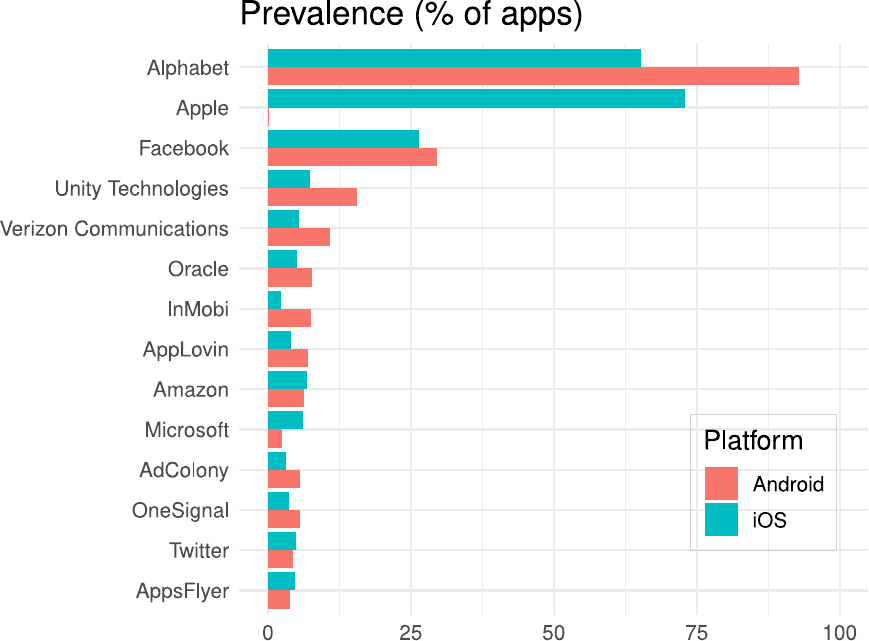}
	    
	    \vspace{0.1cm}
	    
	    \resizebox{\linewidth}{!}{%
	    \begin{tabular}{l r r r r r r}
			& Median
			& Mean
			& Q1
			& Q3
			& Count $>10$
			& None \\
			\midrule
			Android & 1 & 2.3 & 1 & 2 & 2.77\% & 6.86\% \\
			iOS     & 2 & 2.6 & 1 & 3 & 1.82\% & 15.28\% \\
		\end{tabular}}
	\caption{Root companies that are ultimately behind tracking.}~\label{fig:numCompaniesReferred}
\end{figure}

\subsection{Tracker Companies}
\label{sec:tracking_companies}

\textbf{Owners of Tracking Technology.}
Since many tracker companies belong to a larger consortium of companies (see Figure~\ref{fig:verizon} for the example of Verizon), we now consider what parent companies ultimately own the tracking technology, i.e. the \textit{root companies} behind tracker companies.
We report these root companies from combining the observations from our static and traffic analysis, and checking against our X-Ray 2020 (see Section~\ref{sec:company_resolution}).

Figure~\ref{fig:numCompaniesReferred} shows both the prevalence of root parents (i.e. their share among all apps), as well as descriptive statistics.
The median number of companies was 1 on Android, 2 on iOS.
This reflects the fact that Google is prominent in data collection from apps on both platforms, but Apple only on iOS.
The maximum number of companies was 21 on Android, and 23 on iOS.

The overwhelming share of apps share data with one or more tracker companies ultimately owned by Alphabet, the parent company of Google, as can be seen from Figure~\ref{fig:numCompaniesReferred}.
This company can collect data from nearly 100\% of Android apps, and has its tracking libraries integrated into them.
Apple can collect tracking data
(mainly about users' interactions with in-app ads and purchases)
from more than two thirds of apps.
Next most common is Facebook, which has a similar presence on both platforms, and slightly more so on Android.
Tracker companies owned by Unity and Verizon can be contacted by roughly twice as many Android apps than iOS ones.
Beyond these larger companies, a range of smaller specialised tracker companies (including InMobi, AppLovin, AdColony) engage in smartphone tracking. These can potentially pose unexpected privacy risks, since they attract much less scrutiny by regulators and the interested public.

\begin{figure}
    \centering
    \includegraphics[width=0.8\columnwidth]{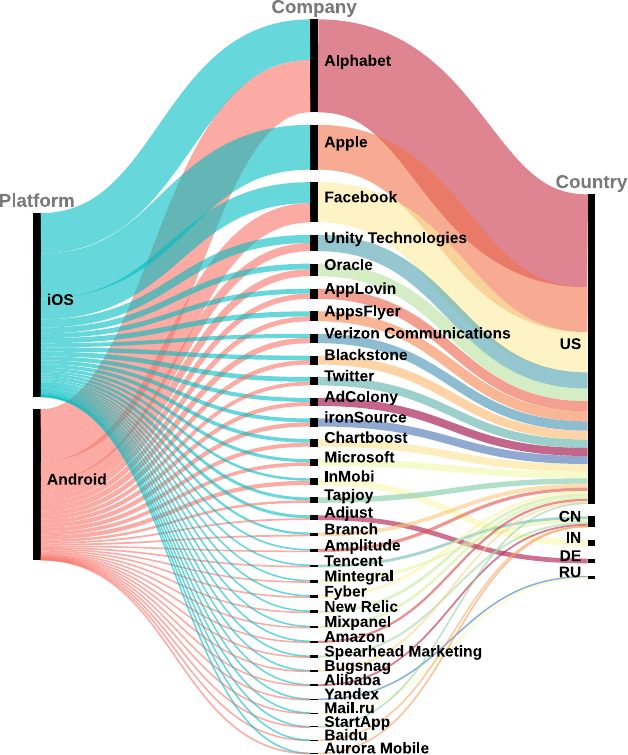}
    \caption{Visualisation of third-party tracking across platforms, root companies, and the jurisdictions of these root companies.}
    \label{fig:alluvial}
\end{figure}

\textbf{Countries of Tracker Companies.}
Based upon the X-Ray 2020 database, which contains company jurisdictions, we now can analyse in what countries the companies behind app tracking are based (including both subsidiary and root parent).
This is visualised in Figure~\ref{fig:alluvial} (root parents only).

The US is the most prominent jurisdiction for tracker companies. 93.3\% of Android apps and 83.5\% of iOS apps can send data to a US-based company.
The next most common destinations are China on iOS (9.5\% of iOS apps; 4.8\% on Android) and India on Android (7.45\% of iOS apps; 2.23\% on Android).
These destinations highlight the global distribution models of both the Apple and Google ecosystem.
While Google Play has a large user base in India, it is not available in China where instead numerous other app stores compete~\cite{china_2018}.
Conversely, the Apple App Store is available in China, and is the only authorised app marketplace on iOS.
Germany and Russia are the only other countries whose root tracker companies reach more than 2\% of apps  on iOS or Android.

While we downloaded apps from the UK app store, the most commonly contacted tracking countries are based outside the UK and EU.
This can give rise to potential violations of EU and UK data protection law, since the exchange of personal data beyond the EU / UK is only legal if special safeguards are put in place, or an \textit{adequacy decision} by the European Commission (or its UK equivalent) exists~\cite{binns2014,binns_third_2018}.
However, such adequacy decisions do not exist for the three most common jurisdictions of tracker companies, namely the US, China and India.
According to our data, the exchange of data with companies based in countries without an adequacy decision seems similarly widespread on both Android and iOS.

\begin{table*}
    \centering
    \resizebox{\textwidth}{!}{
    \begin{tabular}{lllllll}
    Platform & Android &  &  & iOS &  &  \\
    Category & All & Cross-Platform & Children & All & Cross-Platform & Children \\ \midrule
    Total Number & 11\,876 & 1\,623 & 371 & 11\,964 & 1\,534 & 109 \\
    Root Tracker companies & {2.3} & 2.8 & 2.7 & 2.6 & \textbf{3.3} & 2.4 \\
    Permissions (Cross-Pltf.) & 11.0 (0.8) & 14.3 (1.2) & 6.9  ({0.2}) & 3.7  (2.0) & 4.0  (\textbf{2.1}) & 2.7  (1.4) \\
    Location Permission & 28.0\% & 41.1\% & {3.8\%} & 49.2\% & \textbf{53.1\%} & 26.6\% \\
    AdId access (in traffic) & 86.1\% (55.4\%) & 84.4\% (\textbf{64.3\%}) & \textbf{89.8\%} (59.3\%) & {42.7\%} (30.9\%) & 49.9\% (38.3\%) & 50.5\% ({24.8\%}) \\
    \end{tabular}}
    \caption{Comparative statistics for all, cross-platform and children's apps on iOS and Android. Since iOS and Android have permissions of different kinds and absolute numbers, we also provide means both for all and cross-platform permissions (as defined in Section~\ref{sec:code-analysis-manifest}). Column-wise maxima in bold.}
    \label{tab:privacy_comparison}
\end{table*}

\subsection{Cross-Platform Apps}
\label{sec:cross-platform-apps}

Many previous studies pursuing cross-platform app analysis (i.e. analysing both Android and iOS apps) focused on those apps that exist on both platforms.
However, there has been limited discussion of how the characteristics of those \textit{cross-platform apps} might differ from those of the average app on either platform.
Our data suggests notable differences.

Table~\ref{tab:privacy_comparison} shows a comparison between cross-platform and all apps across a range of privacy indicators (and also children's apps, which are discussed in the next Section).
All privacy indicators show worse properties than for all apps: data sharing with tracker companies, the presence of permissions, potential access to location, and the communication of the AdId over the Internet were increased among cross-platform apps.
On Android, cross-platform apps could share data with 2.8 companies on average, compared to 2.3 in the total Android sample.
The difference of $0.5$ in the average number of companies is statistically significant ($p < 0.001$, permutation test with 10,000 permutations).
Their iOS counterparts could share with 3.3 companies on average (compared to 2.6 in the total iOS sample, $p < 0.001$).
The mean number of permissions was increased from 11.0 to 14.3 on Android ($p<0.001$), and from 3.7 to 4.0 on iOS ($p=0.003$).
When focusing on cross-platform permissions, the figure was increased from 0.8 to 1.2 on Android ($p<0.001$), and from 2.0 to 2.1 on iOS ($p=0.007$).
Apps had a similar level of AdId access on Android than across all apps ($p=0.06$). However, more apps (64.3\% in cross-platform apps compared to 55.4\%) were observed to share the AdId over the Internet ($p<0.001$).
Similarly, the proportion of apps that share the AdId over the Internet was increased from 30.9\% to 38.3\% on iOS ($p<0.001$).

The reason for the higher amount of tracking in cross-platform apps may be due to increased popularity, and thereby heightened financial interest in data collection for advertising and analytics purposes.
This makes it not only more valuable to use user data for advertising and other purposes, but also to develop an app for both platforms in the first place.
Indeed, manual analysis showed that among the top 100 apps from the UK app stores on Android and iOS, 92\% existed for both platforms.
The more popular an app,
the more likely it seems to be available on both platforms and the more likely it is to use a greater number of tracking services.

\subsection{Apps for Children}
\label{sec:kids-apps}

Children enjoy special protections under data protection laws in many jurisdictions, including COPPA in the US and the GDPR in the EU and UK.
Among other legal requirements, US, EU and UK law require parental consent for many data collection activities involving children.
The UK's Age appropriate 
design code explicitly prohibits the use of profiling technologies without prior consent~\cite{ico_kids}.
In addition to the legal requirements, Apple and Google impose contractual obligations on children's apps on their app stores.
As such, the study of children's apps not only allows us to assess the practices of apps aimed at particularly vulnerable users, but also serves as a useful case study for the efficacy of privacy rules imposed by policy makers and app platforms.
Both app stores offer a dedicated section for children apps, known as the \textit{Kids} category on the Apple App Store and the \textit{Designed for Families} program on the Google Play Store.
109 iOS apps (0.9\%) and 371 Android apps (3.1\%) from our dataset fell into these categories.
While this dataset is much smaller than in the previous sections, our analysis of this subset suggests that worrying privacy practices are not absent from children's apps, see Table~\ref{tab:privacy_comparison}.

\textbf{Tracking.} On average, tracking~--~in terms of the root companies present~--~was more widespread in Android apps for children than across all apps ($p=0.02$, using a permutation test with $10,000$ permutations and the difference in mean as our test statistic), but not so for iOS ($p=0.41$, 95\% CI $[2.05, 2.76]$).
Most of this tracking is related to analytics purposes on iOS.
84.4\% of iOS apps contained Apple's SKAdNetwork (compared to 69.9\% across all iOS apps), which is used for ad attribution.
The next most common tracking libraries in children's apps on iOS are Google Firebase Analytics (40.4\%, compared to 54.7\% on Android), Google Crashlytics (22.0\%, compared to 14.0\% on Android), and the Facebook SDK (13.8\%, compared to 17.8\% on Android).
As for Android, the most commonly contacted domain (50.4\% of apps) was \texttt{googleads.g.doubleclick.net}, which belongs to Google's advertising business.
71.7\% of Android children's apps contained Google AdMob (compared to 14.7\% on iOS); Unity3d Ads was present in 27.0\% of Android children's apps (compared to 6.42\% on iOS).

\textbf{AdId.} The increased prevalence of advertising-related tracking in children's apps on Android is consistent with the fact that more children's apps on Android were observed to share the AdId over the Internet compared to all apps (59.3\% compared to 55.5\%, $p=0.14$, 95\% CI $[54.3, 64.3]$), but not so on iOS (24.8\% compared to 30.9\%, $p=0.17$, 95\% CI $[16.4, 33.0]$)~--~these observed differences were not statistically significant, but the 95\% confidence intervals still point to common sharing of this identifier over the Internet.
The differences in AdId access between the platforms, and potentially the lower proportions of children's apps on the App Store
might stem from a differing stringency of privacy rules on the two app stores. Apple started to restrict third-party data collection from children's apps~\cite{apple_kids1} from June 2019 onwards. Children's apps \enquote{may not send personally identifiable information or device information to third parties}~\cite{apple_kids2}, which includes personalised advertising.
While the Google Play Store also bans personalised ads in children's apps, the sharing of personally identifiable or device information is not expressly prohibited~\cite{google_kids}.

\textbf{Permissions.} Permission use was, on average, lower than across all apps ($p<0.01$), which could hint at improved privacy properties in children's apps. At the same time, more than one quarter of children's apps on iOS (26.6\%, compared to 49.2\% across all apps, $p<0.001$), and 3.8\% (compared to 28.0\% across all apps, $p<0.001$) on Android request location access.
These results reflect the fact that Google Play apps in the Family category are not allowed to access location~\cite{google_kids}.

It remains unclear from our data 1) why a minority of Android apps still declare the location permissions in their app manifest, and 2) whether some apps might obtain user location in other ways, e.g. through side-channels~\cite{reardon_50_2019}.

\textbf{Conclusions.} The study of children's apps revealed that many share data, including unique identifiers, with tracker companies~--~both on Android and iOS.
The sharing of data with advertising services, including unique user identifiers, tended to be more common on Android than on iOS ($p<0.001$).
At the same time, iOS apps contained the location permission seven times more often than their Android counterparts ($p<0.001$), which can lead to unexpected disclosures of GPS data from children.
Data sharing with third parties often takes place without the necessary parental consent, and despite privacy laws and the policies of the platforms.
Not all comparisons between the subset of children's apps and all apps were statistical significant, but even where this was not the case, the reported 95\% confidence intervals still underlined that worrying data practices are common in children's apps.

%% file: 5_discussion_conclusions.tex
\section{Limitations}
\label{sec:limitations}

It is important to highlight certain limitations of our methodology.
We do not cover all apps available in each app store, only a (large) subset of free apps.
Our sampling method relies on the app stores' search functionality, which might be biased differently on each platform.
We excluded apps that were last updated before 2018, assuming that these are not widely used anymore.
The results of our code analysis must be interpreted with care, since not all parts of an app might be invoked in practice~--~an inherent limitation of this type of analysis.
We used off-device network analysis, which may wrongly attribute some communications; we minimised the impact of this by disabling pre-installed apps if possible.
We also used jailbreaking on iOS and rooting on Android to circumvent certificate validation, which might make some apps alter their behaviour.
For network analysis, we used a phone running Android 7, which was somewhat outdated at the time of data collection, but still widely used~\cite{android_marketshare}.
Compared to other research studies, we did not interact with the studied apps, so as to analyse data sharing without user consent.
We also did not analyse interdependent privacy, i.e. how information disclosed from one individual might affect someone else.
In all parts of our analysis, we consider all apps equally,
regardless of popularity~\cite{binns_measuring_2018} and usage time~\cite{van_kleek_x-ray_2018}, both of which can impact user privacy.
Likewise, we treat all tracking domains, libraries and companies equally, though they might pose different risks to users.

\section{Conclusions \& Future Work}
\label{sec:conclusions}

While it has been argued that the choice of smartphone architecture might protect user privacy,
no clear winner between iOS and Android emerges from our analysis.
Data sharing for tracking purposes was common on both platforms.
Android apps tended to share the AdId, which can be used for tracking users across apps, more often than iOS apps ($p<0.001$).
Permissions, that both Apple and Google deem as particularly dangerous and require user opt-in, were more common among iOS apps (although Android also has a greater range of permissions deemed `not dangerous' and do not require opt-in).

\textbf{Compliance Issues.}
Across all studied apps, our study highlights widespread potential infringements of US, EU and UK data protection and privacy laws. Apps widely use third-party tracking without user consent, lack parental consent before sharing PII with third-parties in children's apps, share more data with trackers than necessary, and send personal data to countries without an adequate level of data protection.

A fundamental compliance issue is the lack of transparency around apps' data practices.
Data protection law obliges apps to disclose their data practices adequately (e.g. Article 13 GDPR).
Privacy policies are one way to do this, but are often inadequate~\cite{reidenberg_ambiguity_2016,maps_2019,reyes_wont_2018,okoyomon_ridiculousness_2019}.
At the same time, design decisions by Apple and Google hinder the interested public from independently assessing the privacy practices of apps.
Apple even applies encryption to all iOS apps and widely uses proprietary technologies, thereby driving researchers analysing iOS apps into legal grey areas. 
On Android, Google has banned the installation of root certificates in unmodified versions of Android (which is necessary to assess apps' network communications),
enabled app obfuscation in release builds by default, and has been taking measures against those who modify their Android device with its SafetyNet (even if this is for research purposes).
These new hurdles to app privacy research are in potential conflict with the transparency obligations under data protection and privacy laws.

\textbf{Apple's and Google's Conflict of Interests.}
Since the platforms take a share of any sales through the app stores (up to 30\%), both Apple and Google have a natural interest in creating business opportunities for app publishers, and letting them collect data about users to drive such sales.
Apple's AdId policies might actively encourage certain app monetisation models to its own benefit (Section~\ref{sec:static_adid}).
Our study also underlined the high market share of Google in mobile display advertising, which itself relies on the collection of user data. Google Ads was potentially present in more than half of apps with ads on both iOS and Android (Section~\ref{sec:static_adid}).

The study of children's apps further illustrated the conflict of interests that app platforms face between user privacy and revenues. Both platforms have policies to limit data collection and advertising in children's apps.
Despite this, access to unique device identifiers, specifically the AdId, and the user location was still common in children's apps. 27\% of children's apps on iOS could request the user location, and 4\% on Android.
About 59\% of Android apps shared the AdId with third-parties over the Internet, 25\% on iOS.
This can be used to build fine-grained profiles about children, putting them at risk~\cite{anirudhchi2021}.

As a result of these conflicts of interests, Google's and Apple's business practices are currently being investigated by competition regulators worldwide, including in the US~\cite{house_antitrust,doj_apple_google}, the EU~\cite{eu_appstore}, Germany~\cite{bundeskartellamt_google}, and the UK~\cite{competition_and_markets_authority_online_2020}. 
Indeed, the US Department of Justice is currently investigating potentially anti-competitive and illegal contracts between the two companies~\cite{doj_apple_google}.

\textbf{Suggestions.}
App platforms are well-positioned to protect user privacy~\cite{greene_platform_2018,hoboken2021,holzer_mobile_2011},
but targeted regulation of app platforms remains largely absent~\cite{hoboken2021}.
This stresses the need for increased transparency around apps' practices.
More transparency could also help build trust around the changing takes by platforms on user privacy, including the scanning of users' photo libraries for Child Sexual Abuse Material (CSAM) as recently proposed by Apple~\cite{apple_csam,abelson_bugs_2021}.
In the meantime, transparency around the privacy practices of apps will remain a challenging target to analyse, as will creating accountability for privacy malpractices.
The tools developed in this work seek to
foster discussion on regulatory and transparency issues around app privacy, and
we share all our tools and data publicly to support such work at \url{https://platformcontrol.org/}.

\textbf{Future Work.}
To mitigate privacy concerns around the use of user tracking, Apple has begun imposing stricter privacy rules since the introduction of iOS 14, including the provision of privacy labels on the App Store and a mandatory opt-in to tracking.
We will assess the impact of these policy changes in future work.
Another important field for further study is the development of a cross-platform app instrumentation tool.
Further research is also needed to develop a holistic approach for the assessment of compliance of mobile apps.